\documentclass{article}
\usepackage{ltwol2e}

 \usepackage{epsfig}

\def\Journal#1#2#3#4{{#1} {\bf #2}, #3 (#4)}

\def\NPB{{\em Nucl. Phys.}~B}
\def\PLB{{\em Phys. Lett.}  B}

\def\be{\begin{equation}}
\def\ee{\end{equation}}
\def\bea{\begin{eqnarray}}
\def\eea{\end{eqnarray}}

\def\bu{B^+}
\def\bd{B^0_d} 
\def\bs{B^0_s}
\def\lb{\Lambda_b}
\def\lc{\Lambda_c^+}
\def\bmix{B^0 \mbox{--} \overline{B^0}}

\def\xeb{\langle x_E \rangle_b}
\def\Zbb{Z^0 \rightarrow b\,{\overline b}}

\begin{document}

\pagestyle{empty}

\begin{titlepage}

\begin{flushright}
{\normalsize
SLAC--PUB--7942\\
September 1998\\}
\end{flushright}

\vspace{2.5 cm}

\begin{center}
      {\large\bf $\bu$, $\bd$ AND $b$-BARYON LIFETIMES}
\end{center}

\vspace{2.0 cm}

\begin{center}
        S. Willocq \\
        Stanford Linear Accelerator Center \\
        Stanford University, Stanford, CA 94309, USA \\ ~~\\  ~~\\
        Representing the SLD Collaboration \\
        Stanford Linear Accelerator Center \\
        Stanford University, Stanford, CA 94309, USA \\
\end{center}

\begin{center}

\vspace{15mm}
 {\bf Abstract}
\vspace{5mm}

\end{center}

{\normalsize
\noindent
We review recent $\bu$, $\bd$ and $b$-baryon lifetime measurements
performed by the LEP, SLD and CDF collaborations.
Lifetime ratios of $\tau(\bu) / \tau(\bd) = 1.070 \pm 0.027$
and $\tau(b~\mbox{baryon}) / \tau(\bd) = 0.77 \pm 0.04$ are obtained
using all existing measurements.
The ratio between charged and neutral $B$ meson lifetimes is in good
agreement with theory but the ratio between $b$-baryon and
$B$ meson lifetimes remains somewhat lower than expected.
}

\begin{center}

\vspace{45mm}
{\normalsize\sl
     Presented at the XXIX$^{th}$ International Conference on High Energy
     Physics,
     23-29 July 1998, Vancouver, Canada.}

\vspace{15mm}
{\footnotesize Work supported in part by Department of Energy Contract
                DE--AC03--76SF00515(SLAC).}
\end{center}

\pagestyle{plain}

\end{titlepage}

\title{$\bu$, $\bd$ AND $b$-BARYON LIFETIMES}

\author{S. WILLOCQ}

\address{Stanford Linear Accelerator Center, P.O. Box 4349,
Stanford, CA 94309, USA\\E-mail: willocq@slac.stanford.edu}


\twocolumn[\maketitle\abstracts{
We review recent $\bu$, $\bd$ and $b$-baryon lifetime measurements
performed by the LEP, SLD and CDF collaborations.
Lifetime ratios of $\tau(\bu) / \tau(\bd) = 1.070 \pm 0.027$
and $\tau(b~\mbox{baryon}) / \tau(\bd) = 0.77 \pm 0.04$ are obtained
using all existing measurements.
The ratio between charged and neutral $B$ meson lifetimes is in good
agreement with theory but the ratio between $b$-baryon and
$B$ meson lifetimes remains somewhat lower than expected.
}]

\section{Introduction}

The study of exclusive $b$-hadron lifetimes provides an important test
of our understanding of $b$-hadron decay dynamics.
Lifetimes are especially useful to probe the
strong interaction effects arising from the fact that $b$ quarks
are not free particles but are confined inside hadrons.
In the naive spectator model, the $b$ quarks are treated
as if they were free and one therefore expects
$\tau(\bu) = \tau(\bd) = \tau(\bs) = \tau(\lb)$.
However, this picture does not hold in the case of charm hadrons
for which the lifetimes follow the pattern
$\tau(D^+)\simeq 2.3~\tau(D_s)\simeq 2.5~\tau(D^0)\simeq 5~\tau(\Lambda_c^+)$.
These factors are predicted to scale with the inverse
of the heavy quark mass squared and
the $b$-hadron lifetimes are thus expected to differ by only 10-20\%.
Using the Heavy Quark Expansion, Bigi et al.~\cite{Bigi} predict
$\tau(\bu) / \tau(\bd) = 1 + 0.05\:(f_B/200\:\mbox{MeV})^2$,
where $f_B$ is the $B$-meson decay constant ($f_B = 200 \pm 40$ MeV),
and $\tau(\lb) / \tau(\bd) \simeq 0.9$.
However, Neubert and Sachrajda~\cite{Neubert} argue that a more
theoretically conservative approach yields
$0.8 < \tau(\bu) / \tau(\bd) < 1.2$ and
$0.85 < \tau(\lb) / \tau(\bd) < 1.0$.

Precise knowledge of exclusive $b$-hadron lifetimes is required
for accurate measurements of $|V_{cb}|$ and $B$ mixing, and
is also an important input parameter for $\Zbb$ electroweak measurements.

\vspace*{-1.8pt}   

\section{$\bu$ and $\bd$ Lifetimes}

The LEP, SLD and CDF collaborations have taken advantage of their
precision vertex detectors and of the significant
boost for $b$ hadrons produced in high energy
$e^+ e^-$ and $\bar{p} p$ collisions
to measure exclusive $b$-hadron lifetimes.
Three main analysis techniques have been used to measure $\bu$ and $\bd$
lifetimes.
The first method relies on fully reconstructed $B$ decays
(e.g. $B \to J/\psi\,K$).
This is the ideal method for a lifetime measurement since there is little
or no modelling uncertainty in the $B$ energy and the sample composition.
However, exclusive branching ratios for $B$ decays are typically small
($10^{-4}$ to $10^{-3}$) which severely limits the statistics available
at current facilities.
The second and most utilized method selects semileptonic decays of
the type $B \to D^{(\ast)} l \nu X$, where the $D^{(\ast)}$ meson is
fully reconstructed. Sample composition can be controlled from the data
using the charge correlation between the lepton and the $D^{(\ast)}$ meson.
A complication arises from decays of the type
$B \to D^{\ast\ast} l \nu$ which spoil the $\bu$ and $\bd$ purity of
the respective $\overline{D^0} l^+$ and $D^{\ast -} l^+$ samples,
and whose rates are not well known.

  The CDF collaboration has finalized a study~\cite{CDF_Dlept}
based on the full Run-I data sample and
corresponding to an integrated luminosity of 110 pb$^{-1}$.
A ``$B^+$'' sample consisting of $\overline{D^0} l^+$ pairs is
selected with fully reconstructed $\overline{D^0} \to K^+ \pi^-$
decays.
Similarly, a ``$B^0$'' sample consisting of $D^{\ast -} l^+$ pairs
is selected by reconstructing the decays
$D^{\ast -} \to \overline{D^0} \pi^-$
where $\overline{D^0} \to K^+ \pi^- (\pi^0)$ or
$\overline{D^0} \to K^+ \pi^- \pi^+ \pi^-$.
The $B$ decay vertex is then formed by intersecting the lepton and
$D^{(\ast)}$ trajectories.

  A fit using decay length and momentum information for
the $\overline{D^0} l^+$ and $D^{\ast -} l^+$ samples yields
$\tau(\bu) = 1.637 \pm 0.058 (\mbox{stat}) ^{+0.045}_{-0.043} (\mbox{syst})$ ps,
$\tau(\bd) = 1.474 \pm 0.039 (\mbox{stat}) ^{+0.052}_{-0.051} (\mbox{syst})$ ps, and
$\tau(\bu)/\tau(\bd) = 1.110 \pm 0.056 (\mbox{stat}) ^{+0.033}_{-0.030} (\mbox{syst})$.
Contamination from $B \to D^{\ast\ast} l \nu$ decays is estimated
to be 10-15\% and constitutes the dominant systematic uncertainty
in the lifetime ratio.

  A third method for lifetime measurements relies on inclusive topological
vertexing, pioneered by the DELPHI and SLD collaborations.
Here, the charged particle topology of the decays is reconstructed and
the separation between charged and neutral $b$ hadrons is achieved simply
using the sum of the charges of all tracks associated with a secondary vertex.
This method has the advantage of large statistics but requires good
control in the detailed simulation of $b$ hadron production and decay.

\begin{figure}
\center
  \epsfxsize=7.0cm
  \epsfbox{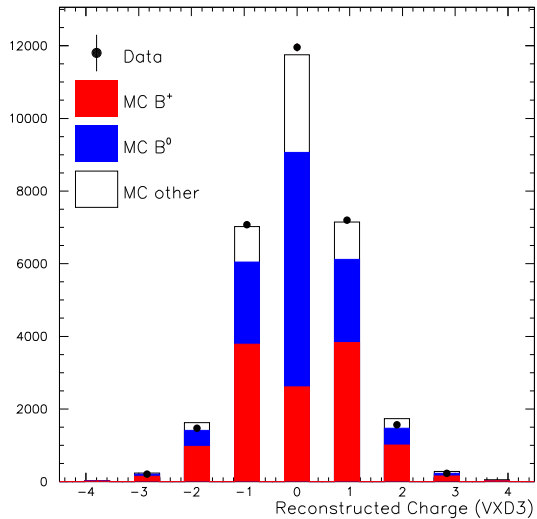}
  \caption{Distribution of the vertex charge for the SLD 1997-98 data
           (points) and Monte Carlo simulation (histograms) indicating the
           contributions from charged and neutral $B$ mesons.
           The category ``MC other'' contains mostly neutral $b$ hadrons:
           $\bs$ and $b$ baryons.}
  \label{fig_SLDqvtx}
\end{figure}

  The SLD collaboration has updated its topological vertexing
analysis~\cite{SLD_topol} with data taken during the first part of the
1997-98 run.
A set of 49,664 $B$ decay candidates is selected with an efficiency
of 50\% and a purity of 98\%.
Separation between $\bu$ and $\bd$ decays is performed on the basis
of the total charge $Q_{tot}$ of tracks associated with the secondary vertex
(see Fig.~\ref{fig_SLDqvtx}).
The charged (neutral) sample consists of 30,028 (19,636) decays
with $|Q_{tot}| = 1, 2, 3$ ($Q_{tot} = 0$).
The charge separation is enhanced somewhat by taking into account the
dependence upon the reconstructed vertex mass
and the $b$-quark charge at production (using techniques developed for
the study of time-dependent $\bmix$ mixing).
An effective $\bu : \bd$ ($\bd : \bu$) separation of $2.6 : 1$
is then obtained in the charged (neutral) sample.

  The lifetimes are extracted with a simultaneous fit to the
decay length distributions of the charged and neutral samples.
Combining with previous data, corresponding to a total sample
of 400,000 hadronic $Z^0$ decays, the lifetimes are
$\tau(\bu) = 1.686 \pm 0.025 (\mbox{stat}) \pm 0.042 (\mbox{syst})$ ps,
$\tau(\bd) = 1.589 \pm 0.026 (\mbox{stat}) \pm 0.055 (\mbox{syst})$ ps, and
$\tau(\bu)/\tau(\bd) = 1.061 ^{+0.031}_{-0.029} (\mbox{stat}) \pm 0.027 (\mbox{syst})$.
These are currently the most precise determinations of the $\bu$
and $\bd$ lifetimes.
The dominant contribution to the lifetime measurement error arises from
the uncertainty in the $b$-fragmentation function. Specifically,
the range of scaled $b$-hadron energy was taken to be
$\xeb = 0.700 \pm 0.011$, which translates into an uncertainty of
$\pm 0.035$ ps in both $\bu$ and $\bd$ lifetimes.
This uncertainty cancels out in the lifetime ratio since all
$b$-hadrons are assumed to have the same fragmentation function.
It should be noted that recent measurements of $\xeb$,
including an analysis by SLD using the same topological
technique,~\cite{SLD_bfrag}
find a somewhat larger value for $\xeb \simeq 0.72$
(see also the L3 measurement below).
Such a value would shift the above lifetimes down by about 0.064 ps.

  The L3 collaboration has also developed an inclusive topological
vertexing technique, first applied to measure the average $b$-hadron
lifetime.~\cite{L3_avg}
The vertexing algorithm uses the 3-D impact parameters and rapidity of
tracks to reconstruct 3 vertices per event corresponding
to the one primary and two secondary vertices expected in $\Zbb$ decays.
Here, the lifetime is extracted from either the secondary vertex decay length
or the impact parameters of tracks attached to the secondary vertex.
The latter has the advantage of having a reduced dependence
on the $b$ fragmentation uncertainty.
Since the two different variables have different sensitivities to this
uncertainty, they can be combined to yield very precise
determinations of both the average $b$-hadron lifetime
$\tau_b = 1.556 \pm 0.010 (\mbox{stat}) \pm 0.017 (\mbox{syst})$ ps
and the average scaled $b$-hadron energy
$\langle x_E \rangle_b = 0.709 \pm 0.004 (\mbox{stat+syst})$.

\begin{figure}
\center
\vskip -0.95cm
  \epsfxsize=7.0cm
  \epsfbox{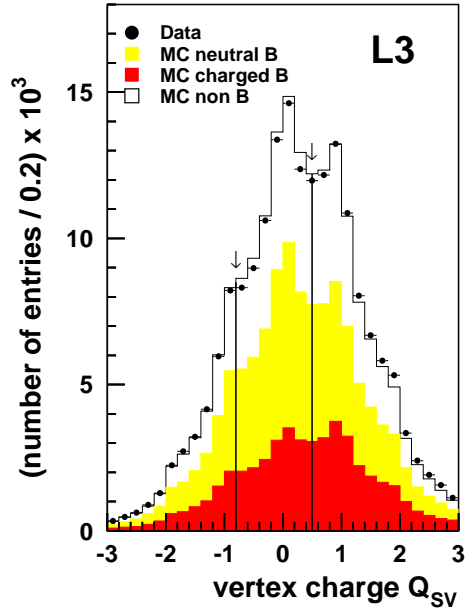}
  \caption{Distribution of the vertex charge for L3 data (points)
           and Monte Carlo simulation (histograms) for charged and neutral
           $B$ mesons.}
  \label{fig_L3qvtx}
\end{figure}

  L3 extended this technique to the study of $\bu$ and $\bd$
lifetimes.~\cite{L3_topol}
From a sample of $2 \times 10^6$ hadronic $Z^0$ decays, the analysis
selects 890,506 secondary vertices.
The separation between charged and neutral decays is then obtained by forming
the vertex charge $Q_{SV}$ defined as the product of the weighted sum of track
charges and the sign of the Jet Charge, where the weight represents the
probability to belong to the secondary vertex.
Fig.~\ref{fig_L3qvtx} shows the vertex charge distribution and the cuts
used to define the charged ($Q_{SV} > 0.5$) and neutral ($-0.8 < Q_{SV} < 0.5$)
samples.
For $Q_{SV} > -0.8$, the sample is 69\% pure in $b$ hadrons.
The $\bu : \bd$ ($\bd : \bu$) separation is estimated to be $1.25 : 1$
($1.10 : 1$) in the charged (neutral) sample.
To reduce the $b$-fragmentation uncertainty, the lifetimes are extracted
using weighted average track impact parameters and a $b$ tag is used in the
opposite hemisphere to suppress the background.
As a result, the lifetimes are found to be
$\tau(\bu) = 1.662 \pm 0.056 (\mbox{stat}) \pm 0.025 (\mbox{syst})$ ps,
$\tau(\bd) = 1.524 \pm 0.055 (\mbox{stat}) \pm 0.037 (\mbox{syst})$ ps, and
$\tau(\bu)/\tau(\bd) = 1.09 \pm 0.07 (\mbox{stat}) \pm 0.03 (\mbox{syst})$.

\begin{figure}
\center
  \hspace*{-6mm}
  \epsfxsize=9.2cm
  \epsfbox{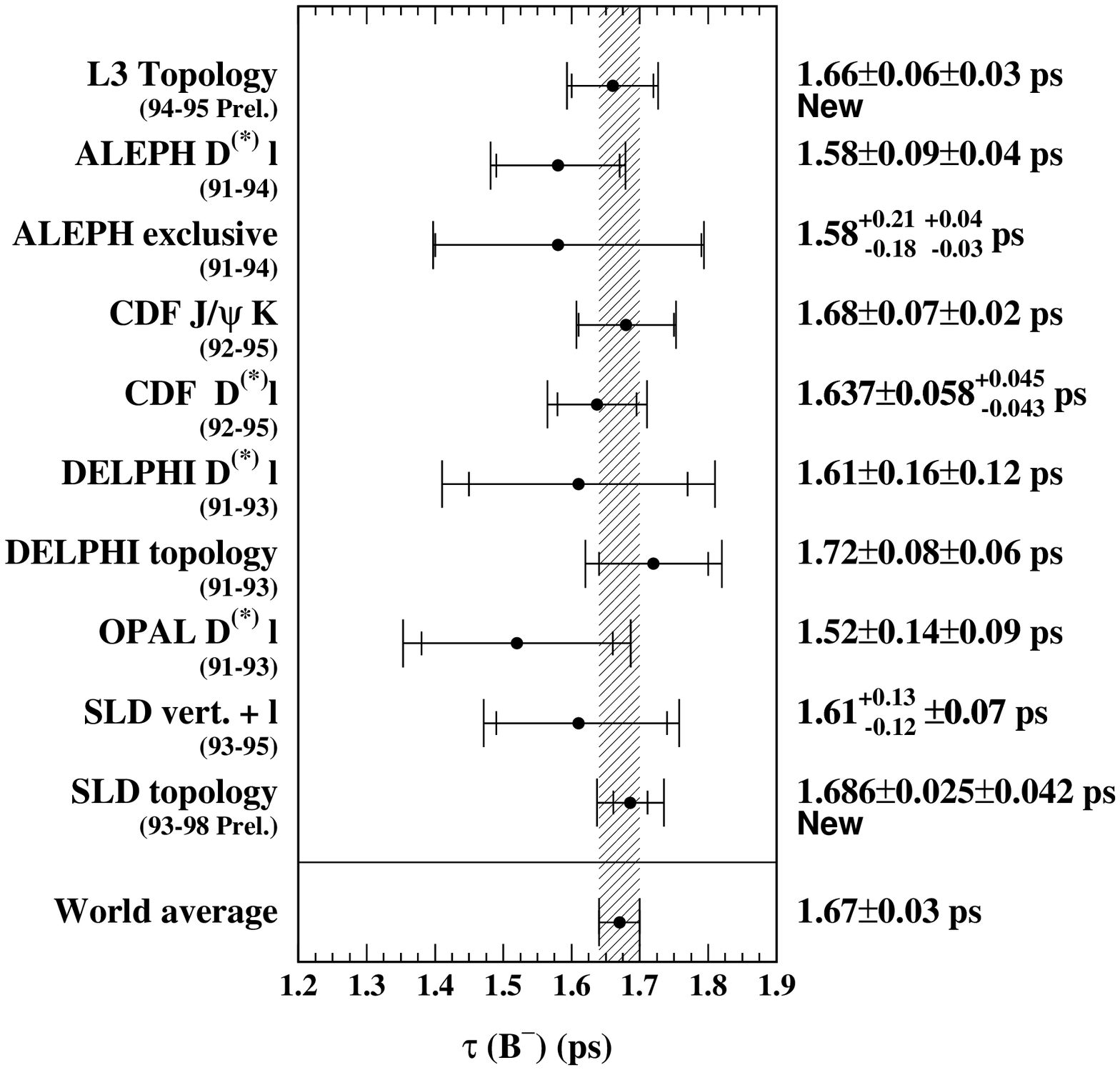}
  \caption{Measurements of the $\bu$ lifetime.}
  \label{fig_bplus}
\end{figure}

\begin{figure}
\center
  \epsfxsize=9.2cm
  \epsfbox{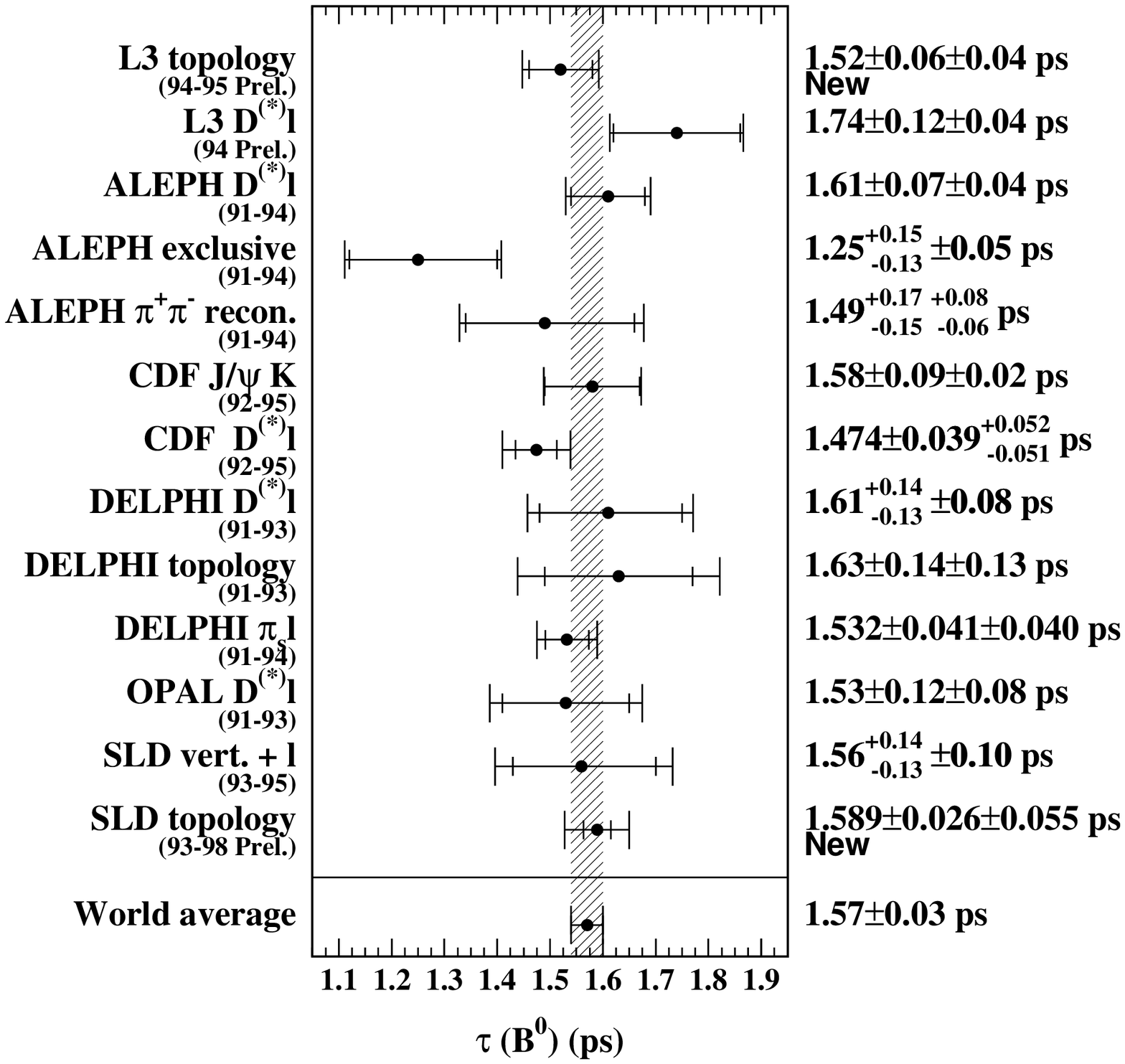}
  \caption{Measurements of the $\bd$ lifetime.}
  \label{fig_b0d}
\end{figure}

\begin{figure}
\center
  \epsfxsize=9.2cm
  \epsfbox{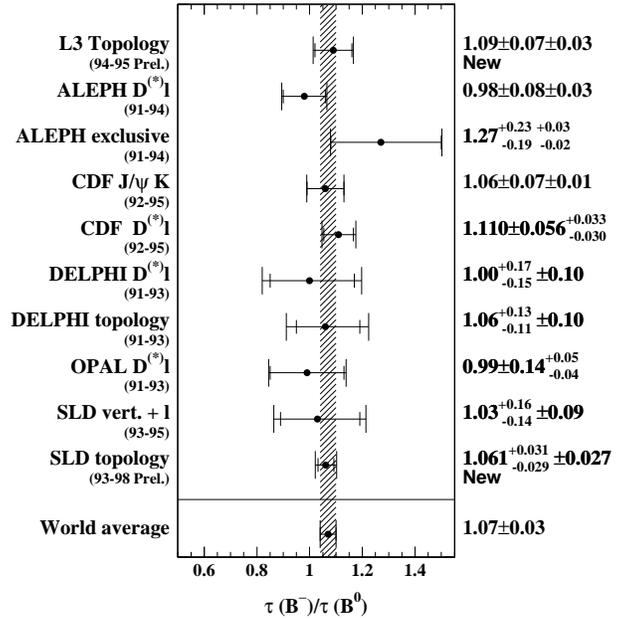}
  \caption{Measurements of the ratio between $\bu$ and $\bd$ lifetimes.}
  \label{fig_ratio}
\end{figure}

The measurements presented above have been combined with all previous
measurements (see Figs. \ref{fig_bplus}-\ref{fig_ratio})
to yield the following world averages:
\bea
  \tau(\bu) & = & 1.67 \pm 0.03 \mbox{~ps}, \\
  \tau(\bd) & = & 1.57 \pm 0.03 \mbox{~ps}, \\
  \tau(\bu)/\tau(\bd) & = & 1.070 \pm 0.027 .
\eea
It is interesting to note that the recent progress in inclusive topological
techniques has allowed a reduction of about 25\% in overall uncertainty since
the last summer conferences. Furthermore, the measurements are becoming
precise enough to begin to measure a difference between $\bu$ and $\bd$
lifetimes.

\section{$b$-baryon Lifetime}

  As mentioned earlier, the lifetime of $b$ baryons is expected to be
about 10\% shorter than that of $\bd$ mesons. However, measurements
over the past few years have indicated that the effect may be as large as
20-25\% which remains somewhat difficult to accommodate.
Measurements of $b$-baryon lifetimes are challenging since $b$~baryons
represent only about 10\% of all $b$ hadrons produced in $\Zbb$ decays
and the properties of $b$ baryons are not well known.
Therefore, most measurements have concentrated on semileptonic decays
and have relied on charge correlations between $\lc$-lepton or
$\Lambda$-lepton pairs to enhance the signal fraction
and control the sample composition.

  The OPAL collaboration has finalized a study~\cite{OPAL_Lb}
of partially reconstructed
$\lb \to \lc l^- \overline{\nu} X$ decays with
$\lc \to p K^- \pi^+$ or $\lc \to \Lambda l^+ \nu X$ decays
in a total sample of $4.4 \times 10^6$ hadronic $Z^0$ events.
The $\lc l^-$ signal is estimated to be $129 \pm 25$ events
and the $\lb$ lifetime extracted from the reconstructed decay length
distribution is
$\tau(\lb) = 1.29 ^{+0.24}_{-0.22} (\mbox{stat}) \pm 0.06 (\mbox{syst})$ ps.

\begin{figure}
\center
  \epsfxsize=8.0cm
  \epsfbox{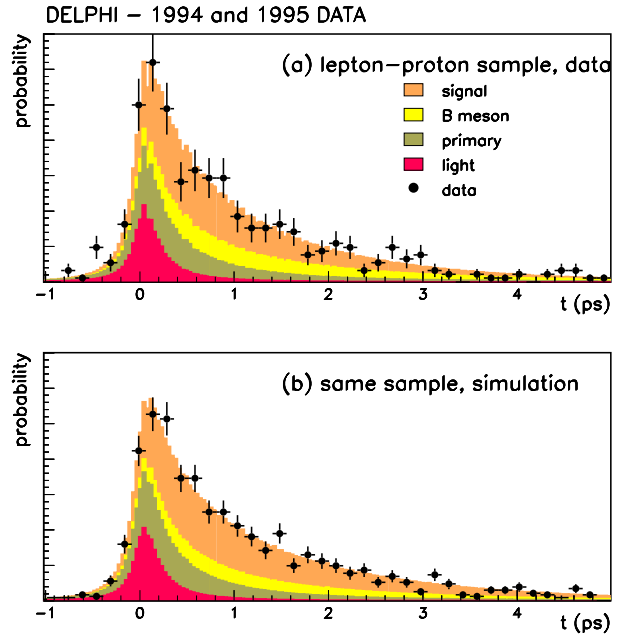}
  \caption{Reconstructed proper time distribution for the DELPHI
           proton-lepton analysis for 1994-95 data (points) and
           the various sample components (histograms).}
  \label{fig_lptime}
\end{figure}

  The DELPHI collaboration released a preliminary study of the same modes
using a sample of $3.6 \times 10^6$ hadronic $Z^0$ events.~\cite{DELPHI_Lb}
Charge-correlations allow the signal fraction to determined from the
data to be $f_{signal} = (56 \pm 6)\%$.
A lifetime fit to the reconstructed proper time distribution yields
$\tau(\lb) = 1.17 ^{+0.20}_{-0.18} (\mbox{stat}) ^{+0.04}_{-0.05} (\mbox{syst})$ ps.
DELPHI also studied more inclusive final states consisting of
$\Lambda$-lepton and proton-lepton pairs.
These have the advantage of increasing the statistical sensitivity of the
measurement but the sample composition is more difficult to control
which leads to higher systematic uncertainties.
The proton-lepton analysis is unique and proceeds by applying an
inclusive reconstruction of $b$-hadron semileptonic decays which relies
on both vertexing and kinematical information.
Then, vertices containing an opposite-sign proton-lepton pair are selected,
where the proton is required to be the fastest hadron in the vertex and
to be positively identified by the RICH particle identification system.
A rejection factor of $~\sim 10$ is achieved for both pion/proton and
kaon/proton separation over most of the momentum range of interest
(3 to 20 GeV/c).
This analysis is only applied to the 1994-95 data sample, corresponding to
$2 \times 10^6$ hadronic $Z^0$ decays, since the RICH was not fully
operational before 1994.
The $b$-baryon lifetime is then extracted from the reconstructed proper
time distribution of the proton-lepton sample (Fig.~\ref{fig_lptime}):
$\tau(b~\mbox{baryon}) = 1.19 \pm 0.14 (\mbox{stat}) \pm 0.07 (\mbox{syst})$ ps
with $f_{signal} = (47 \pm 5)\%$ as estimated from the data.
A study of $\Lambda$-lepton pairs yields
$\tau(b~\mbox{baryon}) = 1.16 \pm 0.20 (\mbox{stat}) \pm 0.09 (\mbox{syst})$ ps
with $f_{signal} = (35 \pm 8)\%$ as estimated from the data.

\begin{figure}
\center
  \epsfxsize=9.2cm
  \epsfbox{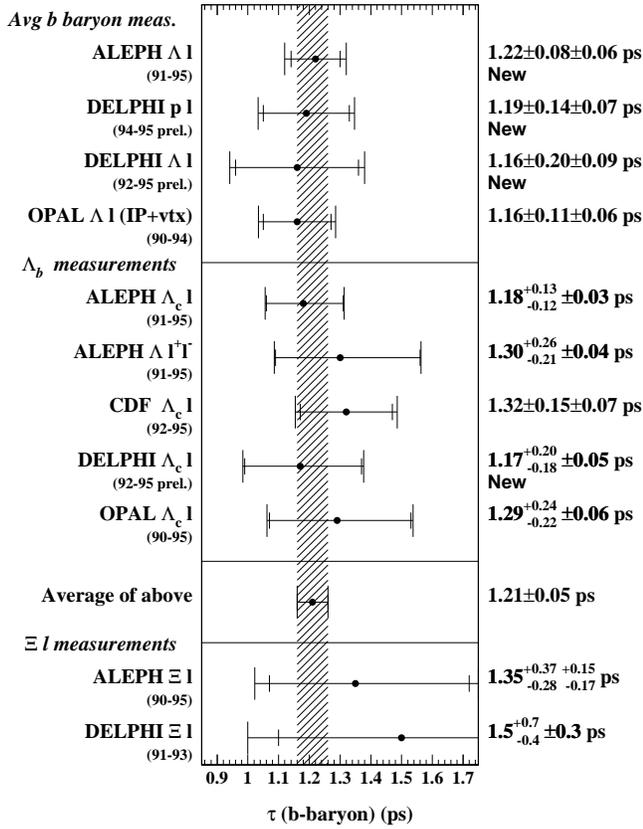}
  \caption{Measurements of the $b$-baryon lifetime.}
  \label{fig_bbar}
\end{figure}

Measurements of the $b$-baryon lifetime are summarized in Fig.~\ref{fig_bbar}.
Averaging $\lc$-lepton with more inclusive $\Lambda$-lepton and 
proton-lepton measurements yields the following world average:
\bea
  \tau(b~\mbox{baryon}) & = & 1.21 \pm 0.05 \mbox{~ps}.
\eea

\section{Summary}

\begin{figure}
\center
\vskip -1.1cm
  \epsfxsize=9.2cm
  \epsfbox{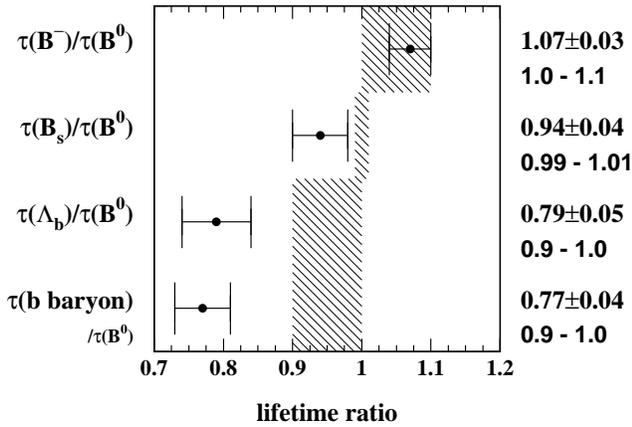}
  \caption{World averages for various $b$-hadron lifetime ratios.
           The hatched bands indicate the approximate range of
           predictions.~$^1$}
  \label{fig_theory}
\end{figure}

$\bu$, $\bd$ and $b$-baryon lifetimes have been measured by the LEP,
SLD and CDF collaborations.
Recent progress in the precision of $\bu$ and $\bd$ lifetimes
has stemmed from the application of inclusive topological techniques
and the addition of new data collected by SLD.
As seen in Fig.~\ref{fig_theory}, lifetime differences are small and
the observed hierarchy
$\tau(\lb) < \tau(\bs) < \tau(\bd) < \tau(\bu)$
is consistent with predictions based on the
Heavy Quark Expansion.~\footnote{A review of $\bs$ lifetime measurements
was presented by A.~Ribon at this conference.}
The measurements are becoming precise enough to begin to see
a difference between $\bu$ and $\bd$ lifetimes, the significance
being at the $2.6\,\sigma$ level.
The $b$-baryon lifetime remains significantly low which
continues to spur theoretical activity.

Further improvements are expected in the near future
from SLD with the inclusion of the full 1997-98 data sample,
corresponding to an increase of $\sim 40\%$ in statistics.
In the longer term, the next step in precision will come from
experiments at the $B$ Factories and the Tevatron.

\section*{Acknowledgments}
I wish to thank Claire Shepherd-Themistocleous, Hans-Gunther Moser and
Juan Alcaraz from the LEP B Lifetime Working Group for updating the
world averages. I have also benefited from interesting discussions
with Claire Bourdarios and Franz Muheim.
John Jaros and Su Dong are thanked for their proofreading.
This work was supported in part by Department of Energy contract
DE--AC03--76SF00515.

\end{document}